\documentstyle[twocolumn,aps,prb]{revtex}

\begin{document}
\title{{\bf Nitrogen local electronic structure in Ga(In)AsN alloys}\\
{\bf by soft-X-ray absorption and emission: Implications for optical
properties }}
\author{V.N. Strocov,$^{1,\ast }$ P.O. Nilsson,$^{2}$ T. Schmitt,$^{3}$ A.
Augustsson,$^{3}$ L. Gridneva,$^{3}$ D. Debowska-Nilsson,$^{2}$ R. Claessen,$%
^{1}$A.Yu. Egorov,$^{4}$ V.M. Ustinov,$^{4}$ Zh.I. Alferov$^{4}$}
\address{$^{1}$Experimentalphysik II, Universit\"{a}t Augsburg,
D-86135Augsburg, Germany}
\address{$^{2}$Department of Physics, Chalmers University of Technology and 
G\"{o}teborg University, SE-41296 G\"{o}teborg, Sweden}
\address{$^{3}$Department of Physics, Uppsala University, \AA ngstr\"{o}m
Laboratory, Box 530, S-75121 Uppsala, Sweden}
\address{$^{4}$A.F. Ioffe Physico-Technical Institute, 194021 St.Petersburg,
Russia }
\date{\today }
\maketitle

\begin{abstract}
Soft-X-ray emission and absorption spectroscopies with their elemental
specificity are used to determine the local electronic structure of N atoms
in Ga(In)AsN diluted semiconductor alloys (N concentrations about 3\%) in
view of applications of such materials in optoelectronics. Deviations of the
N local electronic structure in Ga(In)AsN from the crystalline state in GaN
are dramatic in both valence and conduction bands. In particular, a
depletion of the valence band maximum in the N local charge, taking place at
the N impurities, appears as one of the fundamental origins of reduced
optical efficiency of Ga(In)AsN. Incorporation of In in large concentrations
forms In-rich N local environments such as In$_{4}$N whose the electronic
structure evolves towards improved efficiency. Furthermore, a {\bf k}%
-character of some valence and conduction states, despite the random alloy
nature of Ga(In)AsN, manifests itself in resonant inelastic X-ray scattering.
\end{abstract}


\section{Introduction}

Ga(In)AsN semiconductor alloys are new promising optoelectronic materials,
whose potential applications range from efficient solar cells to laser
diodes operating in the long wavelength range ($\lambda \sim $1.3 $\mu $m)
which fits the transparency window ofoptofibers used in local networks. A
remarkable property of the Ga(In)AsN alloys is an extremely strong
dependence of the band gap width $E_{g}$ on the N content, characterized by
a giant bowing coefficient with $dE_{g}/dx$=15-20 eV (see, e.g., Ref.\cite%
{Kent01}). This figure is more than one order of magnitude larger compared
to the conventional III-V alloys, which suggests that physical mechanisms to
narrow the band gap are quite different. A disadvantage of the Ga(In)AsN
alloys is however their low optical efficiency compared to conventional
III-V alloys such as GaAs and AlAs.

Physics of Ga(In)AsN and related alloys has been under intense study during
the last few years (see Refs.\cite{Kent01,Mattila99,Yacoub00,Yu00,Buyanova01}
and references therein). Due to a strong difference in the N and As
scattering potentials, insertion of N atoms into the host lattice results in
a giant perturbation of the electronic structure and formation of
fundamentally new electronic states such as resonant impurity states. Their
hybridization with the host states in the conduction band strongly perturbs
and shifts these states to lower energies, which narrows the band gap.
Different local environments of N atoms such as isolated impurities, N-N
pairs and various clusters form different states hybridizing with each
other. Because of the immense complexity of such a system no exhaustive
theoretical treatment exists up to now. Different approaches such as the
empirical pseudopotential supercell method,\cite{Kent01,Mattila99}
first-principles pseudopotential method\cite{Yacoub00} and band anticrossing
model\cite{Yu00} often give conflicting predictions. Moreover, their
experimental verification is complicated by significant scatter in the
experimental results depending on the sample preparation. Despite
significant advances in understanding of the band gap narrowing in Ga(In)AsN
alloys, mechanisms responsible for degradation of their optical efficiency
are still not completely clear.

A vast amount of the experimental data on the Ga(In)AsN and similar alloys
has been obtained using optical spectroscopies such as photoluminescence
(PL) and electroreflectance (see, e.g., a compilation of references in Ref.%
\cite{Kent01}). However, they are largely restricted to the band gap region,
and give in general only bare positions of the energy levels without any
direct information about spatial localization or orbital character of
wavefunctions. Such an information can be achieved by soft-X-ray emission
(SXE) and absorption (SXA) spectroscopies with their specificity on the
chemical element and orbital character (see, e.g., a recent review in Ref.%
\cite{Kotani01}). Although their energy resolution, intrinsically limited by
the core hole lifetime, never matches that of the optical spectroscopies,
they give an overall picture of the electronic structure on the energy scale
of the whole valence band (VB) and condiction band (CB). Moreover, as the
orbital selection rules involve the core state and thus engage the VB and CB
states different from those engaged in optical transitions, the soft-X-ray
spectroscopies give a complementary view of the electronic structure.
Because of the small atomic concentrations in diluted alloys such as
Ga(In)AsN, and small crossection of the SXA and SXE processes, these
experiments require the use of 3-generation synchrotron radiation sources,
providing soft X-rays at high intensity and brilliance, and high-resolution
SXE spectrometers with multichannel detection.\cite{Nordgren00}

Extending our pilot work,\cite{Strocov02} we here present experimental SXE
and SXA data on Ga(In)AsN diluted alloys, which unveil the local electronic
structure of N impurities through the whole VB and CB. This yields new
fundamental physics of Ga(In)AsN and related alloys, in particular,
electronic structure origins of their limited optical efficiency, effect of
In on the N\ local environments and electronic structure, and {\bf k}%
-character of some VB and CB states coupled by resonant inelastic X-ray
scattering (RIXS).

\section{Experimental procedure and results}

\subsection{Sample growth}

The Ga(In)AsN samples were grown by molecular beam epitaxy (MBE) at an
EP-1203 machine (Russia) equipped by solid-phase Ga, In and As sources and a
radio-frequency plasma N source. Details of the growth procedure and sample
characterization are given elsewhere.\cite{Egorov01} Briefly, the growth was
performed on a GaAs(001) substrate at 430$^{\circ }$C in As-rich conditions.
The active layer in our GaAsN and GaInAsN samples was grown, respectively,
as GaAs$_{0.97}$N$_{0.03}$ with a thickness of 200 \AA , and In$_{0.07}$Ga$%
_{0.93}$As$_{0.97}$N$_{0.03}$ with a thickness of 240 \AA\ (growth of
thicker layers was hindered by phase segregation). The concentrations of In
and N were checked by high-resolution X-ray rocking curves. A buffer layer
between the substrate and the Ga(In)AsN active layer, and a cap layer on top
of it were grown each as a 50 \AA\ thick AlAs layer sandwiched between two
50 \AA\ thick GaAs layers. Such an insertion of wide band gap AlAs is a
usual method to increase the PL intensity by confining the carriers in the
Ga(In)AsN layer. Moreover, a high-temperature annealing of the grown
structure can be performed after deposition of AlAs in the cap layer without
desorption of GaAs. Such an annealing lasts about 10 minutes at 700-750$%
^{\circ }$C. The resulting improvement of the Ga(In)AsN layer crystal
quality typically increases the PL intensity by a factor of 10-20.

The annealing effect on the local environments of N atoms is less clear. The
N impurities are known to interact with each other due to long-range lattice
relaxation and long tails of their wavefunctions down to N concentrations of
0.1\%,\cite{Kent01} which translates to a characteristic interaction length
of 60 \AA . We expect that on this length scale the annealing can promote
energetically favourable N local environments. In GaAsN such envoronments
are, for example, (100)-oriented N pairs.\cite{Kent01} In the GaInAsN
quaternary alloy the situation is more complicated: whereas as-grown samples
have nearly random distributions of In and N atoms with a significant
fraction of InAs clusters having small chemical bond energy, the annealing
should promote formation of In-N bonds, providing better lattice match to
the GaAs substrate and thus mimimizing the strain energy.\cite{Kim01} In any
case, on a length scale larger than the N interaction length the annealing
should improve homogeneity of the N concentration. This is of paramount
importance, in particular, for our experiment because due to the giant
bowing coefficient of Ga(In)AsN any fluctuations of the N concentration
should result in significant fluctuations of the electronic structure\cite%
{Mintairov01,Matsuda01} and therefore in smearing of spectral structures.

\subsection{SXE/SXA measurements}

The SXE/SXA experiments were performed in MAX-lab, Sweden, at the undulator
beamline I511-3 equipped with a modified SX-700 plane grating monochromator
and a high-resolution Rowland-mount grazing incidence spectrometer.\cite%
{Nordgren89} SXE/SXA measurements employed the N 1$s$ core level at
approximately 400 eV.

The SXA spectra were recorded in the fluorescence yield (FY), because due to
the thick cap layer the electron yield did not show any N 1$s$ absorption
structure. The measurements were performed in partial FY using the SXE
spectrometer operated slitless. It was adjusted at a photon energy window
centered at the N $K$-emission line and covering an interval, in the 1st
order of diffraction, from some 320 to 470 eV. The signal was detected with
the spectrometer position-sensitive detector as the integral fluorescence
within this energy window. Interestingly, usual measurements in the total FY
(detected with a microchannel plate detector in front of the sample)
returned considerably different spectra. This is possibly because the total
FY is more susceptible to irrelevant contributions due to higher-order
incident light and low-energy photoelectron bremsstrahlung fluorescence,
significant with our low N concentrations in the host material. As the
partial FY measurements are characterized by significant intensity loss due
to smaller acceptance angle of the spectrometer, we operated the
monochromator at an energy resolution of 0.45 eV FWHM (the N 1$s$ lifetime
broadening is about 0.1 eV\cite{Lawniczak00}).

The synchrotron radiation excited SXE spectra were measured, in view of the
low crossection of the SXE process and small N concentration, with the
monochromator resolution lowered to $\sim $1.5 eV and to $\sim $0.5 eV for
the off-resonance and resonant spectra, respectively. The spectrometer was
operated with a spherical grating of 5 m radius and 400 lines/mm groove
density in the 1st order of diffraction, providing a resolution of $\sim $%
1.2 eV. The signal from the position-sensitive detector was aberration
corrected using 3rd-order polynomial fitting and normalized to the total
illuminated area in each channel on the detector. Normal data acquisition
time was 2-5 hours per spectrum. Despite the cap layer we could also see a N
signal under 3.5 keV electron beam excitation, although on top of strong
bremsstrahlung background, but this was not suitable for resonant
measurements.

Energy calibration of the spectrometer was performed in absolute photon
energies employing the Ni $L_{l}$, $L_{\alpha _{1,2}}$ and $L_{\beta _{1}}$
lines seen in the 2nd order of diffraction. Abberations in the dispersion
direction of the position-sensitive detector were corrected by setting an
energy scale as a function of the channel number using 2nd order polynomial
fitting. Based on the elastic peaks in SXE spectra, the monochromator was
then calibrated in the same absolute energy scale with an accuracy about $%
\pm $0.15 eV.

\subsection{Experimental results}

Our experimental N 1$s$ SXA spectra (measured in the partial FY)\cite%
{TotalFY} and off-resonant SXE spectra (excitation energy of 420 eV, well
above the absorption threshold) of the GaAs$_{0.97}$N$_{0.03}$ and Ga$_{0.93}
$In$_{0.07}$As$_{0.97}$N$_{0.03}$ samples are shown in Fig.1 ({\it upper
panel}). The binding energy scale is set relative to the VB maximum (VBM)
determined, roughly, by linear extrapolation of the SXE spectral leading
edge. We intentionally give the spectra without denoising to facilitate
judgement the significance of the spectral structures compared to the noise
level. Recent supercell calculations by Persson and Zunger\cite{Persson03}
are in good agreement with our experimental results.

Recent SXA data on GaAs$_{0.97}$N$_{0.03}$ by Lordi {\it et al},\cite%
{Lordi03} which appeared after initial submission of this paper, are
consistent with our results (apart from some energy shift which is
presumably because their energy scale was affected by the monochromator
calibration). Previous SXA data by Soo {\it et al}\cite{Soo99} suffer from
worse experimental resolution and sample quality.

Local environments of the N atoms in Ga(In)AsN are polymorphic,
corresponding to isolated impurities and various clusters.\cite{Kent01}
Applying random statistics, the concentration ratio of the pair and
higher-order N clusters to the total number of N atoms is given by 1-(1-$x$)$%
^{m}$, where $x$ is the N concentration and $m$=4 the number of the nearest
anions in the zinc-blende lattice. With our N concentrations of 3\% this
ratio is only 11.5\%. Therefore, our SXE/SXA spectra characterize mainly the
isolated N impurities.

\section{Discussion}

\subsection{Overall picture of the electronic structure}

The experimental SXA and off-resonance SXE spectra in Fig.1 reflect, by the
dipole selection rules requiring that the orbital quantum number $l$ is
changed by $\pm 1$, the $p$-component of the DOS locally in the N core
region. The $p$-component, by analogy with crystalline GaN,\cite{Lawniczak00}
should in fact dominate the total DOS through the whole VB and CB region.
Core excitonic effects are presumably less significant because the direct
recombination peak\cite{Agui99} does not show up in our SXE spectra.
Splitting of the VBM into the light and heavy hole subbands due to a strain
imposed by the GaAs substrate,\cite{Zhang00} being about a few tenths of eV,
is below our experimental resolution.

It is instructive to compare our Ga(In)AsN spectra to the corresponding
spectra of crystalline GaN. They are reproduced in Fig.1 ({\it lower panel})
in the binding energy scale determined in the same way as for Ga(In)AsN. The
spectra of GaN in the metastable zinc-blende structure, which has the same N
coordination as Ga(In)AsN, were measured by Lawniczak {\it et al},\cite%
{Lawniczak00} and those of wurtzite GaN by Stagarescu {\it et al}.\cite%
{Stagarescu96} Apart from the CB shift, the spectra of the two crystalline
forms are similar in overall shape. They are well understood in terms of the
local orbital-projected DOS and band structure.\cite{Lawniczak00,Lambrecht97}

Comparison of the SXE/SXA data on Ga(In)AsN to those on the two GaN
crystalline structures shows:

(1) In the VB, the overall shape of the SXE signal for Ga(In)AsN is similar
to crystalline GaN. However, the spectral maximum is strongly shifted
towards the VB interior, with the leading edge at the VBM being much less
steep (which has important implications for optical efficiency, see below).
This is not a resolution effect, because the reference spectra of
crystalline GaN were taken at close resolution figures (around 0.8 eV for
zinc-blende GaN and 1.1 eV for wurtzite GaN). Our experimental data
demonstrate thus that the VB electronic structure undergoes, contrary to the
common point of view, significant changes upon incorporation of N atoms into
GaAs. Interestingly, our SXE spectrum did not show any structure due to
hybridization with the Ga 3{\it d} states at $\sim $19 eV below the VBM,
found in wurtzite GaN;\cite{Stagarescu96,Duda98}

(2) In the CB, the differences are radical. The leading peak of the SXA
spectrum for Ga(In)AsN rises immediately at the CB minimum (CBM) and has
much larger amplitude compared to the leading shoulder-like structure in the
spectra of crystalline GaN. The energy separation between the VB and CB
states for Ga(In)AsN is much smaller, which correlates with smaller
fundamental band gap.

On the whole, the observed differences of the Ga(In)AsN spectra to
crystalline GaN manifest that the local electronic structure of the N atoms
in the Ga(In)AsN random alloy is radically different from that in the
regular GaN lattice.

Although further theoretical analysis is required to interpret our
experimental data in detail, we can tentatively assign the leading SXA peak
to the $t_{2}(L_{1c})$ derived perturbed host state which, according to the
calculations by Kent and Zunger on GaAsN,\cite{Kent01} has the strongest N
localization in the CBM\ region. This assignment is corroborated by the
resonant SXE data (see below) which reveals the $L$-character of the leading
SXA peak.

It should be noted that the dipole selection rules in SXE/SXA, inherently
involving transitions from and to the core level, project out the states
from the VB and CB, which can differ from those projected out by the optical
transitions between the VB and CB states themselves. For example,
delocalized states can give only a small contribution to the SXE/SXA signal
due to relatively small overlap with the core state, but they can strongly
overlap with each other and give a strong PL signal. Our SXA data give
explicit examples of this: The $a_{1}(\Gamma _{1c})$ derived states near the
CBM (see Ref.\cite{Kent01}) are not seen in the SXA spectrum due to the
weaker N localization compared to the $t_{2}(L_{1c})$ states, but in optical
spectroscopies they manifest themselves as the intense $E^{-}$ transitions.
On the other hand, the $t_{2}(L_{1c})$ states are not seen in the optical
spectra due to unfavorable matrix elements, but show up as a prominent SXA
peak. Moreover, the energy separation between the VB and CB states in the
SXE/SXA spectra gives only an upper estimate for the fundamental band gap,
because weakly localized N states as well as Ga and As derived states are
not seen. Therefore, the SXE/SXA spectroscopies give a view of the VB and CB
complementary to that by optical spectroscopies.

\subsection{Charge depletion in the VBM: Origin of reduced optical efficiency%
}

The vast body of optical spectroscopy data on Ga(In)AsN evidences that the
optical efficiency sharply drops upon incorporation of the smallest N
concentrations into GaAs, and then decreases further with increase of the N
molar fraction (see, e.g., a compilation in Ref. \cite{Buyanova01}). This is
most pronounced for GaAsN, where the PL intensity loss towards N
concentrations of 5\% is at least 50 times as compared to GaAs. Exact
origins of such a dramatic efficiency degradation are not completely clear.
Supercell calculations in Ref.\cite{Bellaiche97} suggest that about 30\% of
the GaAs efficiency is lost due to gradual smearing of VBM and CBM in their $%
\Gamma $-character, which results in reduction of the optical transition
matrix element. However, this effect is by far weaker compared to the
experimental degradation. Another known origin is relatively poor structural
quality of Ga(In)AsN layers epitaxially grown on GaAs. This is due to,
firstly, low growth temperatures which are used with large N concentrations
to promote high N uptake and, secondly, some lattice mismatch between
Ga(In)AsN and GaAs. However, the first problem can be alleviated by
post-growth high-temperature annealing, and the second by tuning the In
concentration in GaInAsN which allows matching the GaAs lattice constant.
Althought the PL intensity from lattice-matched GaInAsN layers does increase
by a factor about 5 compared to GaAsN, this still remains by far low
compared to GaAs. Moreover, the structural quality does not explain the
efficiency drop at the smallest N concentrations.

Our SXE/SXA results unveil another origin of the optical efficiency
degradation in the very electronic structure. By virtue of the N
localization of the CBM wavefunction\cite{Kent01} the N\ impurities act as
the main recombination centers in Ga(In)AsN. At the same time, the local
valence charge at the N impurities is shifted off the VBM. This appears
immediately from comparison of our experimental SXE spectra of Ga(In)AsN
with those of crystalline GaN, which are in fact representative of GaAs by
virtue of qualitatively similar valence DOS of these materials\cite%
{Chelikowski89} (direct measurements on GaAs are hindered by very low
fluorescence yield of As in the soft-X-ray region). Such a charge depletion
in the VBM, equivalent to reduction of the VBM wavefunction amplitude,
results in a weak overlap of the CBM and VBM wavefunctions at the N\
impurities, which immediately reduces efficiency of the N impurities as
radiative recombination centers. This VBM depletion effect, characteristic
of isolated N impurities, is one of fundamental origins of the reduced
optical efficiency of Ga(In)AsN. Being in play already at the smallest N
concentrations, it immediately explains the initial efficiency drop, whereas
further efficiency degradation with increase of N concentration is
presumably through the structural quality effects.

To explain the observed VBM charge depletion, in Ref.\cite{Strocov02} we
suggested a VBM charge transfer off the N atoms in Ga(In)AsN compared to GaN
(in Ref.\cite{Persson03} this our statement was misinterpreted as a charge
transfer to As, but N has larger electronegativity). In fact, the charge
transfer is more likely to take place not in space but in energy towards
deeper valence states, which is supported by recent computational analysis
of Persson and Zunger.\cite{Persson03} Physically, the local electronic
structure of the N impurities in the GaAs lattice appears somewhere in
between that of the crystalline state,\ and that of isolated atoms. The
observed DOS peaked near the VB center can therefore be viewed as a
transitional case between the DOS of extended band states piling up near the
band edges, and the singularity-like DOS of isolated atoms at the VB center.

Formation of N local environments different from the isolated impurities can
be suggested as a way to increase the optical efficiency of Ga(In)AsN. For
example, in multiatomic N local environments such as clusters of
Ga-separated N atoms the wavefunctions may become closer to crystalline GaN
with its DOS piling up at the VBM. Based on the random statistics, the
cluster concentration should increase with the total N concentration.
Alternatively, the N local environments can be changed by replacing some
neighbour Ga atoms by different cations.

\subsection{Effect of In}

Quaternary GaInAsN alloys, where some Ga atoms are replaced by In, allow
improvement of the optical efficiency by a factor about 5. This is
predominantly due to two factors: a better lattice match of GaInAsN layers
to the GaAs substrate, which improves their structural quality, and electron
confinement effects connected with concentration fluctuations.\cite%
{Mintairov01} We here endevoured investigation whether the incorporation of
In also causes any favourable changes in the electronic structure.

At relatively low In concentrations, evolution of the N local electronic
structure is evidenced by comparison of the Ga$_{0.93}$In$_{0.07}$As$_{0.97}$%
N$_{0.03}$ and GaAs$_{0.97}$N$_{0.03}$ experimental spectra in Fig.1. The
SXE spectra zoomed in the VB\ region are also shown in Fig.2. Surprisingly,
the comparison shows no notable changes within the experimental statistics,
nor in the spectral shapes, neither in energies of the spectral structures.
This evidences that despite the high-temperature annealing the N atoms
reside mostly in In-depleted local environments such as Ga$_{4}$N and
possibly\cite{Kurtz01} In$_{1}$Ga$_{3}$N where the presence of only one In
atom in 4 nearest neighbours should not change the N local electronic
structure dramatically. This experimental finding seriously questions
results of recent Monte Carlo simulations\cite{Kim01} which predict
predominance of In-rich N local environments such as Ga$_{1}$In$_{3}$N and In%
$_{4}$N, at least with low In concentrations. Any effects connected with
insufficient annealing of our samples can be ruled out, as evidenced by
stabilization of PL spectra already after 5 min of annealing. The absence of
any significant electronic structure changes in Ga$_{0.93}$In$_{0.07}$As$%
_{0.97}$N$_{0.03}$ compared to GaAs$_{0.97}$N$_{0.03}$ suggests that at low
In concentrations the optical efficiency improvement is exclusively due to
the structural and electron confinement effects.

To force formation of In-rich N local environments, we have grown a sample
of Ga$_{0.69}$In$_{0.31}$As$_{0.98}$N$_{0.02}$ (170 \AA\ thick active layer)
where the In/N concentration ratio is much increased (the decrease in bare N
concentration is presumably less important because interaction of the
isolated N impurities in such diluted alloys should be weak). The
experimental SXE spectrum of Ga$_{0.69}$In$_{0.31}$As$_{0.98}$N$_{0.02}$,
measured under the same off-resonance conditions as in Fig.1, is also shown
in Fig.2. Now the spectral maximum is shifted by some 0.25 eV to higher
energies compared to GaAs$_{0.97}$N$_{0.03}$, indicating changes in the N
local electronic structure caused by In-rich N environments. Interestingly,
the XAS data for Ga$_{0.7}$In$_{0.3}$As$_{0.97}$N$_{0.03}$ from Ref.\cite%
{Lordi03} demonstrates the CBM simultaneously shifts to lower energies.

The observed VB and CB shifts towards each other suggest that the In-rich N
environments become the main recombination centers in GaInAsN. Moreover,
both holes and electrons become confined in In-reach regions formed by
statistical fluctuations of In concentration on $\mu m$-scale. This effect
increases the optical efficiency of GaInAsN.\cite{Mintairov01}

To see whether the observed changes in the VB affect the optical efficiency
within the above VBM\ depletion mechanism, we examined closely the VBM
region (insert in Fig.2). The shift of the spectral maximum is definitely
larger than that of the VBM (although its exact location requires better
statistics). This indicates certain charge accumulation at the VBM compared
to GaAs$_{0.97}$N$_{0.03}$, and thus increase of the optical efficiency of
In-rich N local environments compared to Ga$_{4}$N. The observed
accumulation seems though rather subtle to explain the increase in GaInAsN
wholly, and the most of it still resides with the structural and electron
confinement effects.

\subsection{{\bf k}-conservation in the RIXS process}

Resonant phenomena were investigated on the GaAsN prototype alloy. Fig.3
shows resonant SXE spectra measured with excitation energies near the two
dominant SXA structures in Fig.1 compared to that measured well above the
absorption threshold. The spectra are normalized to the integral excitation
flux, which was registered from the photocurrent at a gold mesh inserted
after the refocussing mirror.

Intriguingly, not only does the intensity of the resonant spectra increase
in this RIXS process, but also the shoulder at the VB bottom scales up and
becomes a distinct narrow peak at a binding energy of $\sim $7.4 eV. Such a
behavior reveals states near the VB bottom which effectively overlap with
states near the CB\ bottom into which the core electron is excited.

Despite the random alloy nature of Ga(In)AsN, the observed effect can be
interpreted in terms of momentum conservation which appears in the RIXS
process due to coupling of absorption and emission in one single event (see,
e.g., Refs.\cite{Kotani01,Eisebitt98,Carlisle99} and references therein). At
first glance, this should not occur in a random alloy, because the very
concept of momentum, strictly speaking, collapses due to the lack of
translational invariance. However, the description in terms of wavevectors 
{\bf k} can be revived using a spectral decomposition%
\[
\psi ^{N}({\bf r})=\sum\limits_{{\bf k}}C_{{\bf k}}\phi _{{\bf k}}^{GaAs}(%
{\bf r}) 
\]%
of the N-localized wavefunction $\psi ^{N}({\bf r})$ over the Bloch waves $%
\phi _{{\bf k}}^{GaAs}({\bf r})$ of the unperturbed GaAs lattice, each
having a well-defined {\bf k}.\cite{Kent01,Wang98} Then the first SXA peak
is due to the $t_{2}(L_{1c})$ state, whose decomposition is dominated by 
{\bf k} from the $L$-point in the Brillouin zone of GaAs.\cite{Kent01} The
VB bottom, by analogy with the zinc-blende GaN band structure,\cite{Lewis01}
should be dominated by the same $L$-point. The RIXS process will then couple
these points in the CB and VB, blowing up the SXE signal in the VB bottom as
observed in the experiment. Our resonant data demonstrate thus, to our
knowledge for the first time, a possibility for the {\bf k}-conserving RIXS
phenomenon in random alloys.

\section{Conclusion}

Local electronic structure of N atoms in Ga(In)AsN diluted semiconductor
alloys (N concentrations about 3\%) has been determined using SXE/SXA
spectroscopies with their elemental specificity. The experimental N 1$s$
off-resonance SXE spectra and SXA spectra yield the local $p$-DOS of N
impurities through the whole VB\ and CB, complementing information about the
band gap region achieved by optical spectroscopies. The experimental results
demonstrate dramatic differences of the N\ local electronic structure in
Ga(In)AsN from that in the crystalline GaN state. A few peculiarities have
immediate implications for optical properties: (1) The N impurities are
characterized by depletion of the the local charge in the VBM due to charge
transfer towards deeper valence states, which reduces overlap with the CBM
states. This is one of the fundamental origins of the reduced optical
efficiency of Ga(In)AsN. Formation of different N\ local environments can
improve the efficiency; (2) Whereas incorporation of In in small
concentrations has an insignificant effect on the N local electronic
structure, large In concentrations result in formation of In-rich N local
environments whose electronic structure evolves towards improved optical
efficiency. Furthermore, the experimental resonant SXE spectra reveal,
despite the random alloy nature of Ga(In)AsN, a {\bf k}-conserving RIXS
process which couples valence and conduction states having the same $L$%
-character.

\section{Acknowledgements}

We are grateful to A. Zunger and C. Persson for valuable comments and
communicating their computational results before publication. We thank S.
Butorin for his advise on SXE data processing, J. Guo for help with
preliminary experiments, and G. Cirlin for valuable discussions. The work in
the Ioffe institute is supported by the NATO Science for Peace Program
(SfP-972484) and Russian Foundation for Basic Research (project 02-02-17677).

\begin{figure}[tbp]
\caption{({\it upper panel}) Experimental N 1$s$ off-resonant SXE
(excitation energy 420 eV) and SXA spectra of GaAs$_{0.97}$N$_{0.03}$ and Ga$%
_{0.93}$In$_{0.07}$As$_{0.93}$N$_{0.03}$, reflecting the N local $p$-DOS
through the VB and CB; ({\it lower panel}) The corresponding spectra of
crystalline GaN in the zinc-blende \protect\cite{Lawniczak00} and wurtzite
structures \protect\cite{Stagarescu96} shown as a reference. The SXE
spectral maximum for Ga(In)AsN is strongly shifted to lower energies
compared to the crystalline state, resulting in depletion of the N local
charge in the VBM which reduces optical efficiency.}
\end{figure}
\begin{figure}[tbp]
\caption{Experimental off-resonant N1s SXE spectrum of In-rich Ga$_{0.69}$In$%
_{0.31}$As$_{0.98}$N$_{0.02}$ compared to GaAs$_{0.93}$N$_{0.03}$. The
insert details the spectral leading edges (Gaussian smoothed with FWHM of
0.6 eV) which suggest improved optical efficiency of In-rich N local
environments. }
\end{figure}

\begin{figure}[tbp]
\caption{Resonant SXE spectra with the indicated excitation energies
compared to an off-resonant spectrum. The elastic peaks are marked by
vertical ticks. The resonant intensity enhancement in the VB bottom
manifests a {\bf k}-conserving RIXS process. }
\end{figure}


\begin{references}
\bibitem{TotalFY} SXA spectra measured in the total FY (see our Ref.\cite%
{Strocov02}) reproducibly show up a pronounced peak in some 2 eV above the
leading absoption peak, which may be due to structure in the low-energy
bremsstrahlung.
\end{references}
\end{document}